\documentclass[aps,prl,floatfix,longbibliography,twocolumn,footinbib,superscriptaddress]{revtex4-2} 
\usepackage{amsmath}
\usepackage{amssymb}
\usepackage{graphicx}
\usepackage{ulem}
\usepackage{mathtools}
\usepackage{mathrsfs}
\usepackage[colorlinks=true,linkcolor=blue,citecolor=blue,filecolor=blue,urlcolor=blue,pdfstartview=FitH,breaklinks=true]{hyperref}
\usepackage{xcolor}
\usepackage{ulem}

\begin{document}

\title{
Thermodynamics of photonic nonlinear Aharonov-Bohm cages
}

\author{Stefano Iubini}
\affiliation{Istituto dei Sistemi Complessi, Consiglio Nazionale delle Ricerche, via Madonna del Piano 10, I-50019 Sesto Fiorentino, Italy}
\affiliation{Istituto Nazionale di Fisica Nucleare, Sezione di Firenze, via G. Sansone 1 I-50019, Sesto Fiorentino, Italy}

\author{Carlo Danieli}
\affiliation{Istituto dei Sistemi Complessi, Consiglio Nazionale delle Ricerche, via dei Taurini 19, I-00185 Rome, Italy}
\affiliation{Dipartimento di Fisica e Astronomia, Universit\'a di Firenze, via G. Sansone 1 I-50019 Sesto Fiorentino, Italy}

\begin{abstract}

We investigate equilibrium and non-equilibrium thermodynamics of one-dimensional photonic diamond lattices with Kerr nonlinearity.
The equilibrium phase diagram is obtained as a function of the synthetic magnetic flux acting on each plaquette.
In the linear regime, the magnetic flux can induce Aharonov-Bohm caging, flattening all Bloch bands and suppressing particle and energy currents. In this caging regime, non-vanishing currents are enabled by nonlinearity. 
By imposing stationary temperature- and chemical potential- imbalances at the system boundaries, we show that at weak nonlinearity fine tuning the flux at the Aharonov-Bohm caging  transforms the system from a conductor to an insulator. For intermediate nonlinear strength, the system remains conducting for all magnetic fluxes; however, the caging condition significantly enhances the Seebeck coefficient and thermoelectric figure of merit, improving the thermoelectric features of the system.
Our results give evidence of a novel route towards optimization of coupled transport devices, based on the control of
linear versus nonlinear conduction channels via a synthetic magnetic flux.

\end{abstract}

\maketitle

{\it Introduction--} 
in the last decades, systems featuring macroscopically degenerate states gained increasing attention. This interest follows the unconventional phases of the system yielded by perturbations of the degeneracy. 
Flat band lattices (FB) is one of the most notable examples of this class of models~\cite{leykam2018artificial,rhim2021singular,danieli2026progress}. 
FB are spatially periodic tight-binding lattices featuring at least one dispersionless ({\it flat}) Bloch band. 
FB with finite-range hopping host compact localized states (CLS) -- {\it i.e.} they have strictly zero amplitude outside a specific region due to local destructive wave interference. 
Photonics emerged as one of the most prolific platforms for FB research~\cite{leykam2018perspective,vicencio2021photonic,danieli2024flat}. 
The first experimental realizations of CLS in arrays of optical waveguides~\cite{vicencio2015observation,mukherjee2015observation,Xia2016demonstration} indeed paved the way to the theoretical and experimental exploration of {\it e.g.} 
driven FB lattices~\cite{Brosco2021nonabelian,danieli2024thouless,mukherjee2017observation,Sun2022non} and compact solitons in nonlinear FB systems~\cite{vicencio2013discrete,johansson2015compactification,danieli2018compact}. 

Fine-tuning destructive interference can flatten {\it all} bands in a lattice (ABF). This complete lack of dispersion leads to the perfect localization of any initially compact state at any frequency within a finite spatial domain. 
Known as {\it Aharonov-Bohm caging} (AB), this disorder-free localization was first introduced in a 2D Dice lattice dressed with fine-tuned magnetic field~\cite{vidal1998aharonov} and then extended to several flux-dressed 1D chains~\cite{creutz1999end,vidal2000interaction,brosco2021twoflux}. 
Among these setups, the flux-threaded diamond lattice, experimentally realized with photonic waveguide arrays~\cite{mukherjee2018experimental,caceres2022controlled,vicencio2025multi}, serves as an archetypal platform for investigating exotic phases emerging from perturbations to AB caging.
Within the framework of the discrete nonlinear Schr\"odinger equation (DNLS), which is relevant for studying high power light propagation in optical waveguides~\cite{efremidis2002discrete,heinrich2011nonlinear}, it emerged that certain ABF lattices support nonlinear AB caging~\cite{gligoric2019nonlinear,diliberto2019nonlinear,danieli2021nonlinear}, localized states in momentum space~\cite{Chang2021nonlinear} and nematic phases~\cite{kim2026nematic}. 

In this work, we investigate photonic nonlinear AB cages from a thermodynamical perspective -- an approach that has gained significant attention within the field of optics~\cite{wright2022physics}. 
Indeed, several complex phenomena relevant to statistical mechanics have been discovered in highly multimode nonlinear optical systems, including solitons~\cite{Sun2024multimode}, modulation and parametric instabilities~\cite{kraych2019statistical,krupa2016observation}, beam self-cleaning~\cite{Mangini2022statistical,ferraro2023spatial} and mode locking~\cite{wright2017spatio,ferraro2025wave}.  
In recent years, these observations have led to a thermodynamic formalism~\cite{wu2019thermodynamic,Makris2020stat,selim2022thermodynamics,pourbeyram2022direct,dinani2025universal} and an optical kinetic theory~\cite{kurnosov2024optical,kabat2026optical} specifically tailored to nonlinear multimode photonic systems. 
This approach has recently begun reaching the FB research, exploring {\it e.g.} the thermal features of the linear Lieb and Kagome lattices~\cite{wen2025thermal,lara2026inter}, linear AB caging~\cite{pena2026bosonic}, and the thermoelectric efficiency in single AB rings~\cite{mishra2024majorana,garcia2026high}.

\begin{figure}[h!]
    \centering
   \includegraphics[width=0.95\columnwidth]{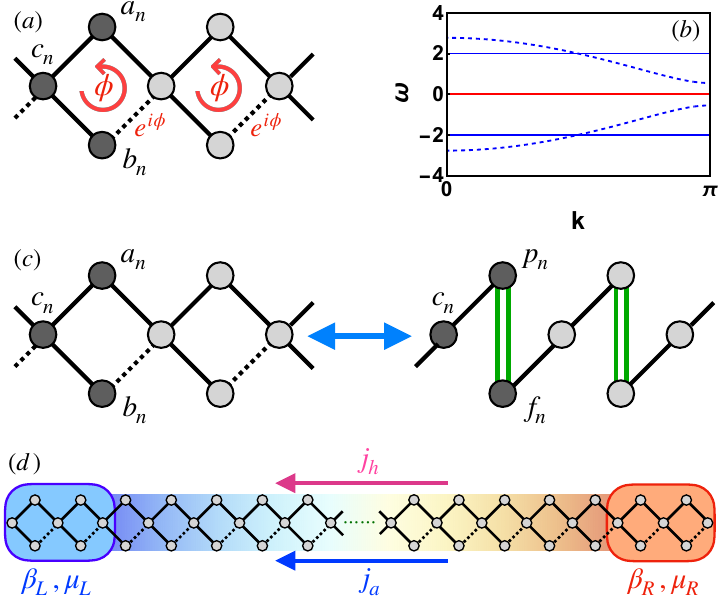}
    \caption{Nonlinear Aharonov-Bohm lattice and non-equilibrium setup. 
    (a) Diamond lattice profile with magnetic field of flux $\phi$. The black dots indicate the fundamental unit-cell. 
    (b) Band structure for $\phi=\frac{\pi}{4}$ (dashed) and $\phi=\pi$ (solid). The $\phi$-independent flat band $\omega_0=0$ is shown in red color. 
    (c) Detangling rotation $U$ of the nonlinear chain for $\phi=\pi$. The black lines represent the linear hopping, while the double green lines represent the nonlinear connections between disconnected trimers. 
    (d) Schematic representation of the MonteCarlo reservoir method, where the left (L) and the right (R) ends of the lattice are connected to a bath (colored in blue and orange respectively) each featuring an inverse temperature $\beta_i$ and a chemical potential $\mu_i$ for $i=L,R$  
    }
    \label{fig:1}
\end{figure}

We consider the diamond lattice threaded with a uniform synthetic magnetic field~\cite{vidal2000interaction} in presence of mean-field cubic nonlinearity~\cite{gligoric2019nonlinear,diliberto2019nonlinear}.
We found that tuning the magnetic field 
strongly dictates the thermal features of the systems. 
First, we demonstrate how the ground state and infinite temperature lines in the norm-energy configuration space vary upon tuning the flux. 
Then, by coupling the chain to imbalanced reservoirs at the edges, we investigate the heat and norm currents versus the flux at different nonlinear strength regimes. 
Nonlinearity indeed establishes pathways for macroscopic currents that would otherwise be suppressed by the linear AB caging~\cite{danieli2021nonlinear}. 
We support these latter findings via Onsager theory of coupled transport, in particular by computing the system's Seebeck coefficient and figure of merit. This reveals that, when the nonlinear and the hopping strengths are comparable, fine-tuning the flux at the AB condition significantly enhances the system's thermoelectric performance.

{\it Model--} we consider a photonic diamond lattice in presence of uniform magnetic field of flux $\phi$ in each rhombic plaquette, and local Kerr nonlinearity. 
This quasi 1D chain is described by Hamiltonian $H=H_0+\frac{\gamma}{2} H_1$
\begin{equation}
\begin{split}
H 
&=\sum_{n=1}^N\left[ a_n^*  (c_n+c_{n+1}) + b_n^* (c_n+ e^{i\phi} c_{n+1})  +  \text{H.c.} \right] \\
&+\frac{\gamma}{2} \sum_{n=1}^N \left[ |a_n|^4 + |b_n|^4 + |c_n|^4 \right] 
\end{split}
\label{eq:Ham}
\end{equation}
where 
$\gamma$ the nonlinear strength, while the hopping is renormalized to one. Here, $N$ is the number of unit-cells, and the chain is considered with open boundary conditions. 
The lattice is shown in Fig.~\ref{fig:1}(a).  The complex variables $a_n, b_n$ and $c_n$ are the time-dependent optical fields, whose dynamics are governed by the Schr\"odinger equations 
\begin{equation}
\begin{split}
i\dot{a}_n &= c_n +c_{n+1}  + \gamma a_n  |a_n|^2\\
i\dot{b}_n &= c_n + e^{i\phi} c_{n+1} + \gamma b_n |b_n|^2 \\
i\dot{c}_n &= a_n +a_{n-1} + b_n + e^{-i\phi} b_{n-1} + \gamma c_n |c_n|^2 
\end{split}
\label{eq:Ham_eq_Nlin}
\end{equation}
Additionally to the Hamiltonian $H$, this system conserves the total norm (power) $A=\sum_{n=1}^N[|a_n|^2+|b_n|^2+|c_n|^2]$.

In the linear regime $\gamma=0$, the lattice is characterized by three Bloch bands 
\begin{equation}
\begin{split}
\omega_0=0 \qquad
\omega_{\pm} = \pm 2\sqrt{1 + \cos\frac{\phi}{2} \cos k}
\end{split}
\label{eq:bands}
\end{equation}
The central band $\omega_0$ is dispersionless for any flux value~\cite{ramachandran2017chiral}, while the upper and lower ones $\omega_{\pm}$ are in general dispersive. 
By fine-tuning the flux to $\phi=\pi$, the dispersive bands $\omega_\pm$ become $k$-independent, and all bands become flat at $\omega_i=0,\pm2$ for $i=0,\pm$, as illustrated in Fig.~\ref{fig:1}(b). The flat band $\omega_0=0$ is shown in red. The two outer bands are dispersive for $\phi=\frac{\pi}{4}$ (shown in blue dashed lines) and flat at $\phi=\pi$ (solid blue).  
Due to the complete lack of dispersion, the system exhibits Aharonov-Bohm caging, where any initially compact state remains therein localized indefinitely in time. 
The unitary transformation 
\begin{equation}
\begin{split}
p_n = \frac{a_n+b_n}{\sqrt{2}}\qquad
f_n = \frac{a_n-b_n}{\sqrt{2}}
\end{split}
\label{eq:transformation}
\end{equation} 
where $c_n$ remain untouched 
recasts the Hamiltonian in Eq.~\eqref{eq:Ham} for $\phi=\pi$ to 
\begin{equation}
\begin{split}
H 
&=\sqrt{2}\sum_{n=1}^N \left[ c_n (p_n^* + f_{n-1}^*) +  \text{H.c.} \right] \\
&+ \frac{\gamma}{4} \sum_{n=1}^N \left[  |p_n|^4 + (p_n^*f_n)^2 + (p_nf_n^*)^2 \right. \\ 
&\qquad\quad \left.+|f_n|^4+4|p_n|^2|f_n|^2 + 2 |c_n|^4 \right] 
\end{split}
\label{eq:Ham_rot1}
\end{equation}
while the equations in Eq.~\eqref{eq:Ham_eq_Nlin} turn to -- see~\cite{Supple} for details 
\begin{equation}
\begin{split}
i\dot{p}_n &=\sqrt{2} c_n + \frac{\gamma}{2} \left[ p_n|p_n|^2 + f_n^2 p_n^*  + 2p_n |f_n|^2     \right]   \\
i\dot{f}_n &= \sqrt{2}c_{n+1} + \frac{\gamma}{2} \left[ f_n|f_n|^2  + p_n^2 f_n^* +  2 |p_n|^2 f_n  \right]  \\
i\dot{c}_n &= \sqrt{2}(p_n +f_{n-1}) + \gamma |c_n|^2 c_n \\
\end{split}
\label{eq:Ham_eq_nonlin1}
\end{equation} 
The transformation~\eqref{eq:transformation}, by removing most of the linear hopping terms, reduces the linear Hamiltonian $H_0$ to a set of disconnected trimers $\{p_n,c_n,f_{n-1}\}$ -- as shown with the black lines in Fig.~\ref{fig:1}(c), right panel. 
On the other hand, the local Kerr nonlinearity becomes nonlocal, connecting neighboring trimers -- as shown with the double green lines. 
For $\gamma=0$, each of the decoupled trimers retains its initial norm and energy. 
For $\gamma\neq 0$ instead, the nonlocal nonlinear terms 
allow the exchange of energy and norm between two neighboring trimers if both are simultaneously non-empty~\footnote{If trimer is initially empty, $p_n(0)=0$, $c_n(0)=0$, and $f_{n-1}(0)=0$, the nonlinear equations of motion Eq.~\eqref{eq:Ham_eq_nonlin1} reduce to $i\dot{p}_n =0$, $i\dot{f}_n =0$, and  $i\dot{c}_n=0$ -- hence, the solution of each of the three complex fields is constant at zero.}.

We exploit this microscopic transport mechanism in a thermodynamics context. 
We consider the nonlinear chain connected to a left and right baths, as shown in Fig.~\ref{fig:1}(d). Each bath is characterized by an inverse temperature $\beta_i$ and a chemical potential $\mu_i$ ($i=L,R$). 
Stationary norm and energy currents $j_a$ and $j_h$ are induced by imbalances at the boundaries. 

\begin{figure}[h!]
    \centering
   \includegraphics[width=0.95\columnwidth]{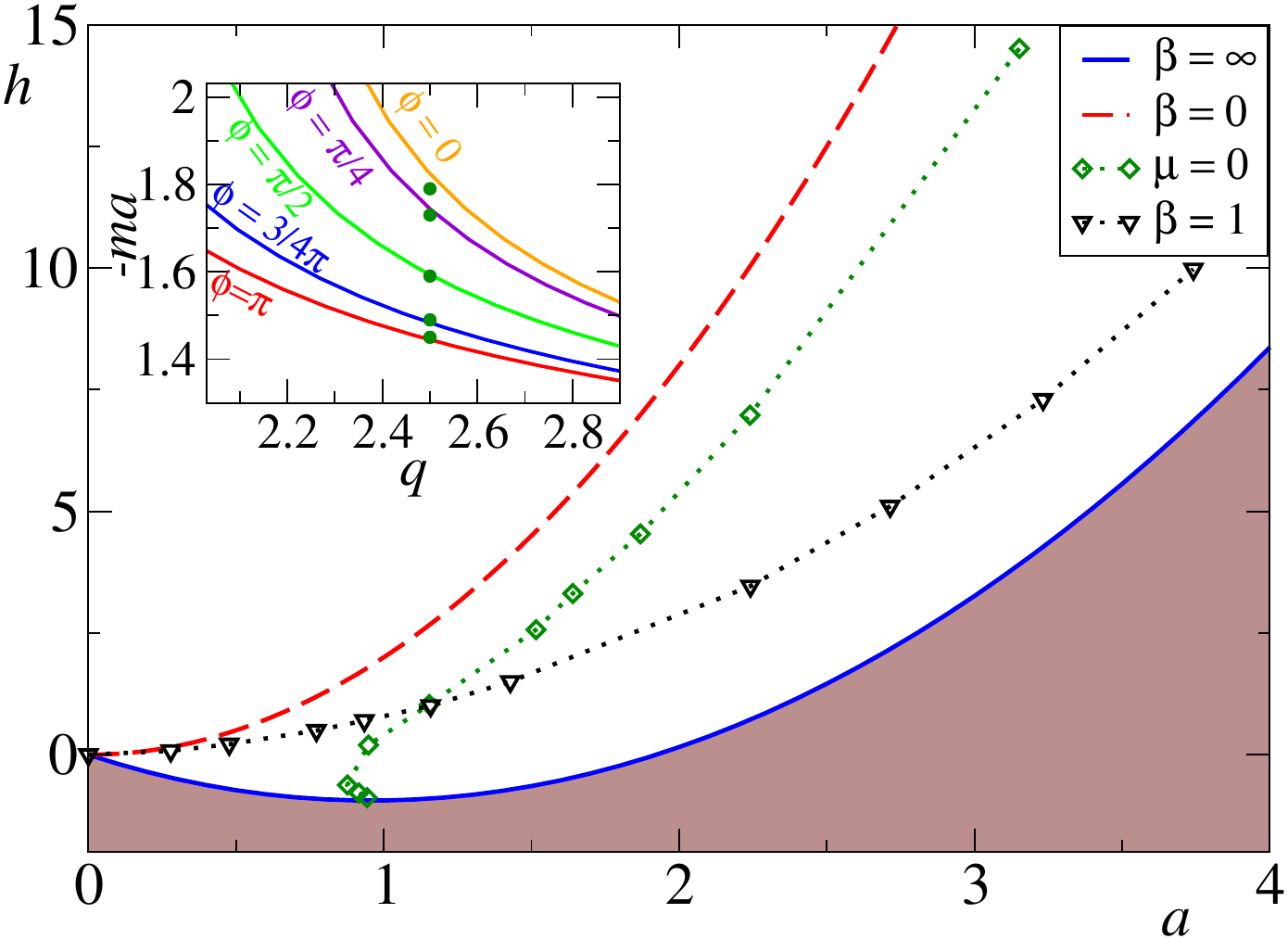}
    \caption{
    Equilibrium phase diagram for $\gamma=2$ and $\phi=\pi$. Symbols refer to equilibrium simulations with $N=20$.
    Inset: harmonic limit.  Rescaled average norm $-ma$ versus $q$ according to Eq.~(\ref{eq:a_gen}) (solid lines).
    Green circles refer to equilibrium numerical simulations (green circles) performed for $q=2.5$ and $m=-1$ for
    corresponding values of $\phi$.
    Simulations  were performed evolving Eq.~(\ref{eq:Ham_eq_Nlin}) in the presence of grand canonical reservoirs. Data shown refer to $N=50$ and time
    averages $t=5\times10^6$ units.
    }
    \label{fig:2}
\end{figure}

{\it Equilibrium--} 
we begin discussing the equilibrium thermodynamics features of the system as function of the flux $\phi$. 
Let us consider the grandcanonical probability distribution $\mathcal{P}(\psi_n)=\frac{1}{Z} \exp[-\beta (H-\mu A)] $ where $\psi_n = (a_n, b_n, c_n)$, and
$Z$ is the partition function. Here $\beta$ and $\mu$ are considered uniform throughout the system. 
We define the average norm density 
$a= \frac{\langle A \rangle }{3N}$ and the average energy density $h= \frac{\langle H \rangle }{3N}$, 
where $\langle \cdot \rangle$ is the average over $\mathcal{P}$. An equilibrium state is identified by a couple $(a,h)$ depending on $(\beta,\mu)$.
Two limiting cases are analytically treatable. 
The $\beta=0$ line is $h=\gamma a^2$, since the contribution of the harmonic part $H_0$ to $h$ is vanishing and phases are uniformly random.
Therefore, the density $h$ is determined solely by the nonlinear terms independently on $\phi$, as in the standard cubic DNLS equation~\cite{rasmussen2000statistical}.
Conversely, the ground-state $\beta=\infty$ line is specific of the diamond lattice geometry and exists for any flux value. For $\phi=\pi$, using the rotated representation in Eq.~(\ref{eq:Ham_rot1}), minimization of $H$ for fixed $A$ leads to the solution of a quartic polynomial in $|c_n|^2$. Here the ground state is characterized by constant
profiles of $|c_n|^2=C$ and $|p_n|^2=|f_n|^2=P$, with $C=3a-2P$ and $P/(3a)$ ranging from $1/4$ $(\gamma=0)$ to $1/3$ $(\gamma\longmapsto\infty)$, see~\cite{Supple}.
The $\beta=\infty$ and $\beta=0$ lines are shown in blue solid and red dashed line respectively in the phase diagram $(a,h)$ in Fig.~\ref{fig:2} for $\gamma=2$ and $\phi=\pi$. The brown shaded area below the $\beta=\infty$ line is not accessible. The area above the $\beta=0$ line is accessible, although the gran canonical distribution $\mathcal{P}$ diverges for large $|\psi_n|$. This region  corresponds to negative-temperature states defined either in the microcanonical 
ensemble~\cite{Gradenigo21} or as grandcanonical metastable states~\cite{iubini25}. Here, spatial localization effects are expected to occur according to
constraint-driven condensation mechanisms~\cite{Nossan14}.

We focus on the region between the $\beta=\infty$ and $\beta=0$ lines, which is homogeneous and positive-temperature.
Examples of isothermal curve $\beta=1$ (purple triangles) and constant chemical potential $\mu=0$ (green diamonds) have been obtained numerically 
by attaching an open Aharonov-Bohm chain to Monte-Carlo reservoirs acting at the boundaries -- see~\cite{Supple} for details.
In the linear regime $\gamma=0$, the distribution $\mathcal{P}$ becomes the well-known Rayleigh-Jeans distribution and $Z =\prod_{k}\prod_{i=0,\pm } \frac{2\pi}{(\beta \omega_i(k) -m)}$ defined for $m=\beta \mu$.
A universal form of equation of state holds in this limit, namely 
\begin{equation}
    a=-\frac{1}{3m}\left( \frac{2q}{\sqrt{K-2q+q^2}}  +1\right) \quad ;  \quad h=\frac{1}{\beta}+\mu a\, ,
    \label{eq:a_gen}
\end{equation}
where  $q=\mu^2/4$ and $K =(1 -\cos \phi)/2$, with $q>1+\sqrt{1-K}$ to ensure real solutions -- see~\cite{Supple} for details.
Eq.~\eqref{eq:a_gen} remarks that the flux $\phi$ has a role in equilibrium thermodynamics already at linear level, as also studied in~\cite{pena2026bosonic} in the context of thermodynamic cycles. 
 The inset of Fig.~\ref{fig:2} illustrates this linear limit. Solid curves are analytic predictions according to Eq.~(\ref{eq:a_gen}) and circles correspond to equilibrium numerical simulations.
Note that the relations in Eq.~\eqref{eq:a_gen} are valid also for $\beta<0$.

{\it Transport--} we now 
study the stationary transport of norm and energy when the system is driven form the boundaries, as shown in Fig.~\ref{fig:1}(d).  Currents of  norm $j_a$ and energy $j_h$ are defined by the average exchanges of norm and energy between the system and the baths over a sufficiently long time~\cite{Supple}. Upon calling $j_a^{(L,R)},j_h^{(L,R)}$ the currents from the Left/Right reservoir to the system, one has $j_a^{(L)}=-j_a^{(R)}\equiv j_a$ and $j_h^{(L)}=-j_h^{(R)}\equiv j_h$ in the stationary state. 

In the linear regime, Landauer theory~\cite{sheng1996} yields the estimates for the norm and heat currents. For the sake of simplicity, here we specialize to the case in which $\mu_L = \mu_R\equiv \mu$.  We obtain
\begin{equation}
\begin{split}
j_a &= C_\phi
\left[\frac{1}{\beta_L} - \frac{1}{\beta_R} \right]\\
j_h &= \left\{ 2(\omega_u - \omega_l) + \mu C_\phi \right\} \left[\frac{1}{\beta_L} - \frac{1}{\beta_R} \right]
\end{split}    
\label{eq:ja_jh_2}     
\end{equation}
for $C_\phi = \log\left[ \frac{(\omega_u-\mu)}{(\omega_u+\mu)} \frac{(\omega_l+\mu)}{(\omega_l-\mu)} \right]$ and the dispersive band edges $\omega_l=2\sqrt{1-\cos{\phi/2}}$ and $\omega_u=2\sqrt{1+\cos{\phi/2}}$, see~\cite{Supple} for details.
\begin{figure}[h!]
    \centering
   \includegraphics[width=0.95\columnwidth]{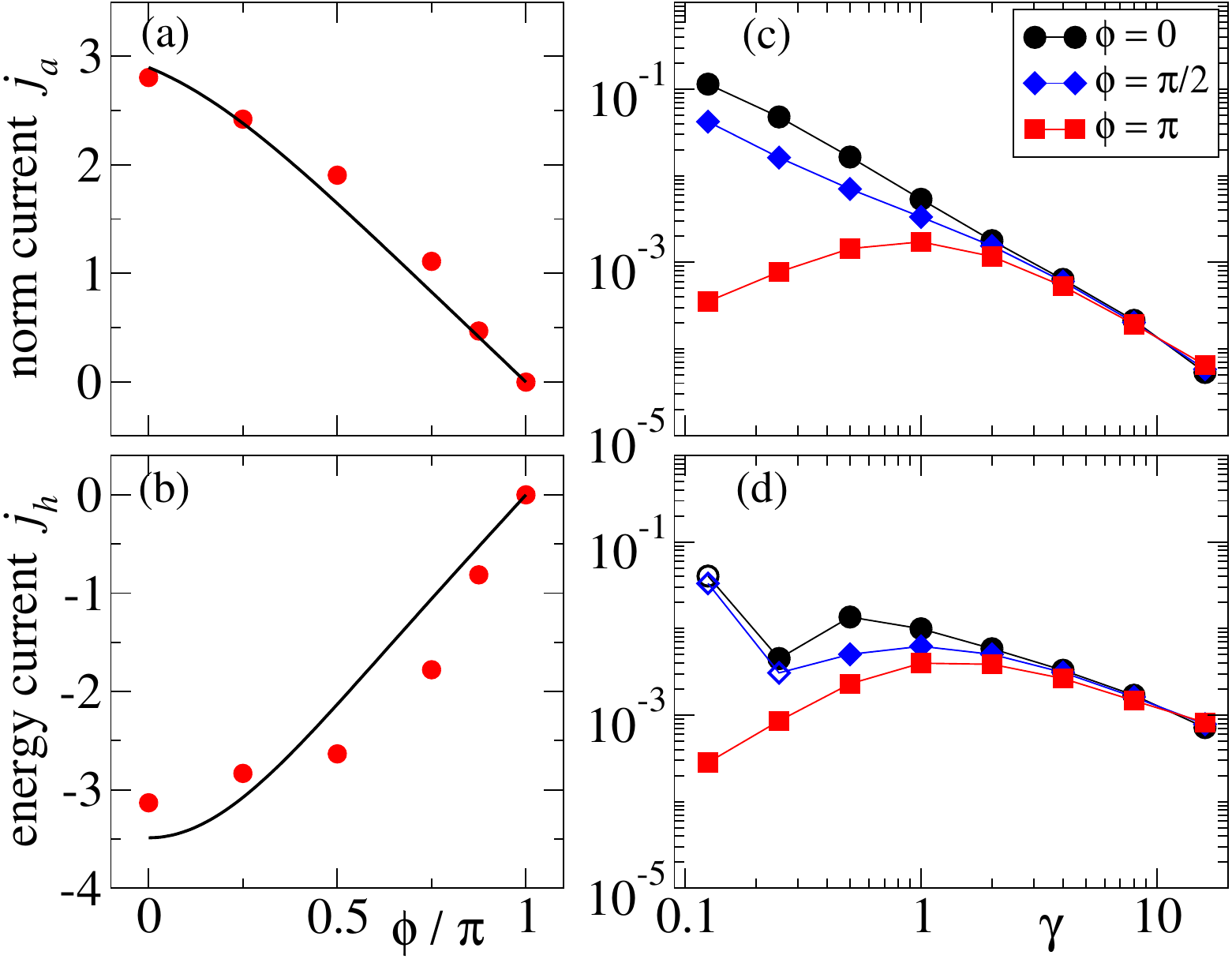}
    \caption{
    Stationary currents of norm (first row) and energy (second row) for $\beta_L=0.2$, $\beta_R=0.316$, $\mu=-3.16$ and $N=50$.  
    (a-b) Dependence of $j_a$ and $j_h$ on the flux $\phi$ in the linear limit: symbols refer to numerical simulations, solid curves are obtained from Landauer theory in Eq.~\eqref{eq:ja_jh_2}.
    Numerical fluxes are rescaled by an arbitrary constant factor.
    (c-d) Dependence of $j_a$ and $j_h$ on the nonlinear parameter $\gamma$ for $\phi=0$ (black), $\phi=\frac{\pi}{2}$ (red) and $  \phi=\pi$ (blue). In (d), positive (negative) values of $j_h$ are shown with full (empty) symbols.
    }
    \label{fig:3}
\end{figure}
In Fig.~\ref{fig:3}(a,b), we show the analytical estimates versus numerical simulations of  $j_a$ and $j_h$ as function of the flux $\phi$ for $\gamma=0$.  
The relatively good agreement confirms the validity of the non-equilibrium setup. Notice that in the limit $\phi\to\pi$ the system becomes an insulator, as both $j_a$ and $j_h$ vanish.
In Fig.~\ref{fig:3}(c,d), we show numerical results of currents $j_a$ and $j_h$ as function of the nonlinear strength $\gamma$ for flux values $\phi=\{0,\frac{\pi}{2},\pi\}$ in black, red and blue respectively.
For $\phi=\{0,\frac{\pi}{2}\}$, both currents decrease monotonically with increasing $\gamma$.  The inflection shown by $j_h$ in Fig.~\ref{fig:3}(d) is due to the fact that  the current changes sign  within the interval $\gamma\leq 1$. 
For $\phi=\pi$ instead, both currents are nonmonotonic in $\gamma$ -- namely they initially increase for increasing $\gamma\lesssim 1$, reaching their maxima for $\gamma\approx 1$ and decrease as $\gamma$ further grows.  
Notice that, for $\gamma \ll 1$ tuning the magnetic field has strong impact on the fluxes $j_a$ and $j_h$ -- {\it e.g.} already for $\gamma=0.1$ the value of the fluxes range over two orders of magnitude between $\phi = 0$ and $\phi=\pi$. 
On the other hand, for $\gamma\gtrsim 5$ both fluxes measured for all three values of $\phi$ are indistinguishable. In particular, as $\gamma\longmapsto \infty$, $j_a$ and $j_h$ vanish since in this regime the lattice in Eq.~\eqref{eq:Ham} decouples into a set of independent oscillators -- {\it i.e.} anti-continuum limit~\cite{MacKay1994proof}. 
We verified that in this strong nonlinear regime, both fluxes $j_a$ and $j_h$ scale as $N^{-1}$ with the system size, thus exhibiting diffusive transport, while they manifest a quasi-ballistic behavior~\cite{lepri2003thermal} in the weak nonlinear regime -- as detailed in~\cite{Supple}.

\begin{figure}[h!]
    \centering
\includegraphics[width=0.95\columnwidth]{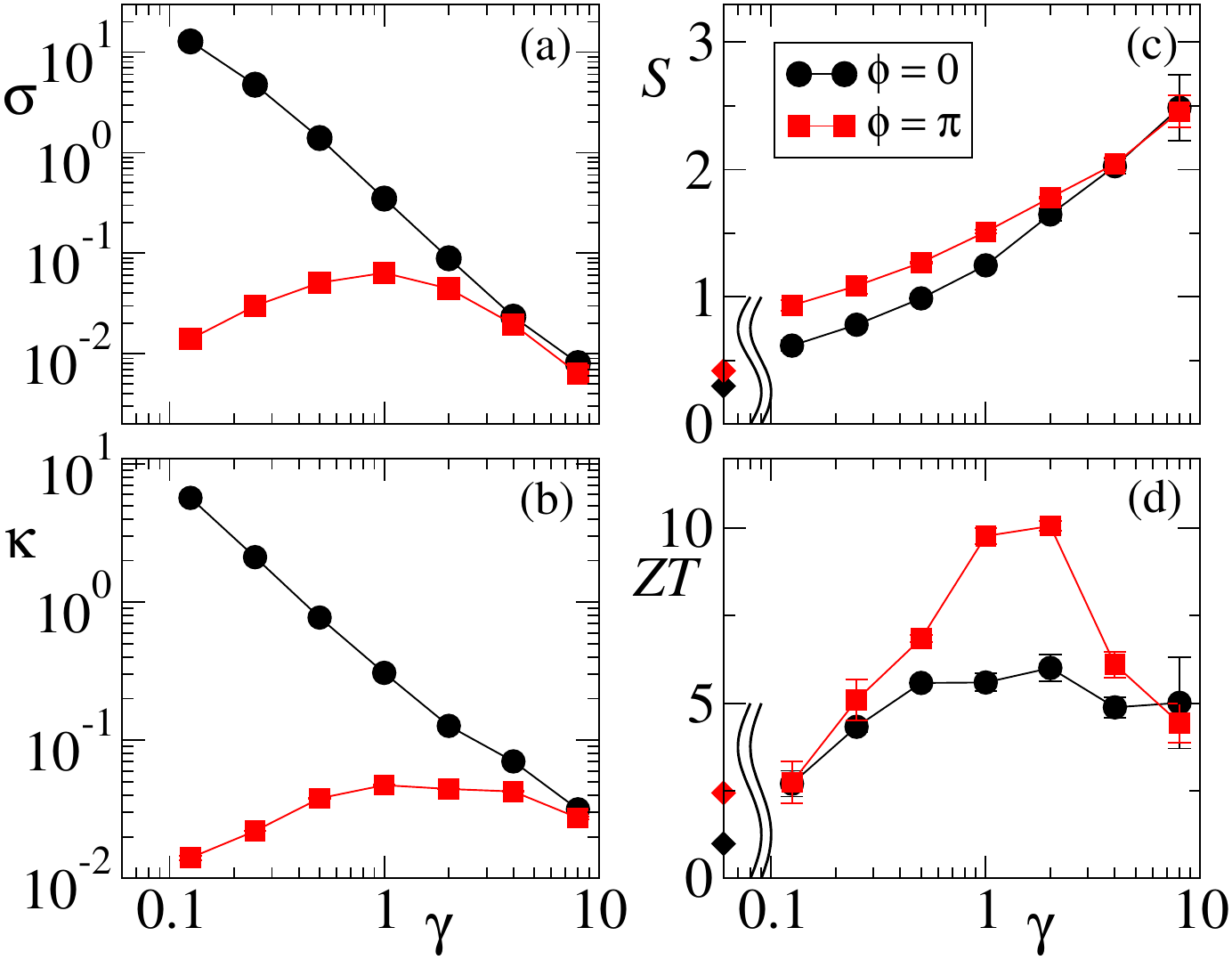}
    \caption{Coupled transport in systems. Norm conductivity $\sigma$ (a), heat conductivity $\kappa$ (b), Seebeck coefficient $S$ (c) and figure of merit $ZT$ (d) versus $\gamma$ computed for $\phi=0$ (black) and $\phi=\pi$ (red) with $N=100$, $\beta=0.32$ and $m=-1$. 
    Diamonds in (c,d), correspond to analytical estimates for $\gamma=0$ via Landauer theory. 
    }
    \label{fig:4}
\end{figure}

For a systematic description of thermoelectric coupled transport of the system, we rely on linear response theory -- specifically on Onsager relations~\cite{iubini12,iubini23,lepri2016thermal,Ioffe1957Semiconductor}
\begin{eqnarray}
\label{eq.Ons1}
j_a &=& -L_{aa} m_y + L_{ah}\beta_y \\
j_h &=& -L_{ha} m_y + L_{hh}\beta_y, 
\label{eq.Ons2}
\end{eqnarray}
which can alternatively be expressed in matrix form~\footnote{Onsager relations in Eqs.~(\ref{eq.Ons1},\ref{eq.Ons2}) in vector form read ${\bf j} = L {\bf f}$, where the $2\times 2$ matrix $L=${\footnotesize$\begin{pmatrix} -L_{aa} & L_{ah} \\ - L_{ha} & L_{hh}\end{pmatrix}$} encapsulates the Onsager coefficients, while ${\bf j} = (j_a,j_h)^t$ and ${\bf f} = (m_y,\beta_y)^t$.}. 
Eqs.~(\ref{eq.Ons1},\ref{eq.Ons2}) relate the currents $j_a$ and $j_b$ to the gradients $m_y = \partial_y m$ and $\beta_y = \partial_y \beta$, where $y = \frac{n}{N}$ is considered in the limit $N\longmapsto \infty$, via the Onsager coefficients $L_{xy}$. 
These coefficients yield the norm conductivity $\sigma=\beta L_{aa}$ and the heat conductivity $\kappa=\beta^2 \det L / L_{aa}$.
Off-diagonal ones bring information on the coupling of currents. We analyze two limit cases corresponding to reversible dynamics: $\phi=\{0,\pi\}$, respectively depicted in Fig.~\ref{fig:4} in black and red symbols. 
In Fig.~\ref{fig:4}(a,b) we show $\sigma$ and $\kappa$ as function of $\gamma$ -- see~\cite{Supple} for details on their computations. 
Notably, both conductivities behave analogously to the currents $j_a$ and $j_h$ shown in Fig.~\ref{fig:3}(c,d) for both values of $\phi$. This remarkable consistency validates the robustness of these two independent measurements.

The Onsager relations in Eqs.~(\ref{eq.Ons1},\ref{eq.Ons2}) also elucidate the synergy between the charge and heat conductivities. This relation is conventionally described by two quantities: the Seebeck coefficient $S= \beta\frac{L_{ah}}{L_{aa}} -m$, which quantifies the heat-to-norm current conversion in analogy with thermoelectricity, 
and the thermoelectric figure of merit  $ZT=\sigma S^2 /(\beta \kappa)$, which measures the material's efficiency at converting thermal energy into optical power~\cite{Ioffe1957Semiconductor,lepri2016thermal}. 
As shown in Fig.~\ref{fig:4}(c), fine-tuning the flux to $\phi=\pi$ consistently enhances the Seebeck coefficient relative to $\phi=0$ within the regime $0\leq \gamma\lesssim3$. Beyond this threshold ({\it i.e.} for $\gamma\gtrsim 3$) adjusting the magnetic flux no longer influences $S$. 
A similar trend emerges from the figure of merit, as shown in Fig.~\ref{fig:4}(d). Remarkably,  tuning the magnetic field from $\phi=0$ to $\phi=\pi$ doubles $ZT$  around $\gamma\approx 1$.   
Some insight on the enhancement of ZT is gained for $\gamma = 0$ from Landauer theory,  see diamonds in Fig.~\ref{fig:4}(c,d). It occurs because the distribution of available transport frequencies collapses onto three discrete points at $\omega=\{0,\pm 2\}$ -- a process known to enhance the thermoelectric efficiency of a material~\cite{mahan1996best}. 
The insulating nature of the system for $\phi=\pi$ for $\gamma=0$ however requires a perturbation of the AB cages (in this case, the Kerr nonlinearity) to activate macroscopic heat and charge currents.

Narrowing of the distribution of the available frequencies towards an ideal width is a highly sought after mechanism to raise $ZT$ of a system, hence improving its thermoelectric efficiency~\cite{mahan1996best,shakouri2005thermoelectric,benenti2011increasing}. 
A common way to achieve this involves the implementation of energy filtering schemes on the transmission coefficient to enhance certain frequencies~\cite{gayaner2020energy,masci2024large,ishibe2025robust,garmroudi2025energy}.
Recent results however show the existence of an {\it optimal} width of frequency for efficient transport in interacting systems~\cite{zhou2011optimal}. 
Our setup provides a new way to control the range of the available frequencies 
by tweaking the system's parameters -- specifically the magnetic flux and the nonlinear strength. As illustrated in Fig.~\ref{fig:1}(c), tuning the flux from $\phi=0$ to $\phi=\pi$ smoothly deforms a nonlinear rhombic chain with bottlenecks at every unit-cell into a 1D chain with a nonlinear hopping every third site. Due to this effect, transport is forced to occur within narrower regions with respect to $\phi=0$.
Our results indicate that the ideal frequency range is achieved for $\gamma\approx 1$, ({\it i.e.} comparable with the normalized linear hopping strength).


{\it Discussion and perspectives--} 
We have studied one of the most notable examples of AB cages with Kerr nonlinearity from a thermodynamical standpoint and revealed that nonlinearity enables non-zero stationary currents, whose features can be controlled via the magnetic field $\phi$. 
In particular, we demonstrated that fine-tuning AB caging 
(i) turns the lattice from a conductor to an insulator for $\gamma\ll1$, and 
(ii) significantly improves the thermoelectric efficiency of the system for $\gamma\approx 1$, while only mildly lowering both heat and norm conductivities by less that one order of magnitude. 
Given the {\it uniform} magnetic field, this effect is expected to be accessible within standard Onsager reciprocal relations in place of Onsager-Casimir ones~\cite{Luo20}.
Our results open a new intriguing venue in the rapidly growing field of optical thermodynamics, {\it i.e.} exploit nonlinear AB cages to engineer materials with enhanced thermoelectric properties, in both one and higher dimensions~\cite{danieli2021nonlinear}. This includes exploring the negative temperature regions of nonlinear AB caging lattices (the area above the red dashed line in Fig.~\ref{fig:2}), where unconventional non-equilibrium and dynamical freezing phenomena emerge already for the 1D DNLS~\cite{iubini2019dynamical}, as well as entering the realm of quantum thermodynamics~\cite{campbell2026roadmap}.

{\it Acknowledgments--} 
We thank  Luca Chirolli, Stefano Lepri and Antonio Politi for useful discussions. 
S.I. acknowledges support from the MUR PRIN2022
project "Breakdown of ergodicity in classical and quantum many-body systems" (BECQuMB) Grant No. 20222BHC9Z. 
C.D. acknowledges support from the Project PNRR MUR No. PE 0000023-NQSTI, Project PNRR MUR Project No. CN 00000013-ICSC, Fondazione Cariplo Grant No. 2023-2594.

{\it Data availability--} 
The data that support the findings of this article are not publicly available. The data are available from the authors upon reasonable request.

\bibliography{ABcage_Th}

\clearpage

\section{Supplemental Material}

\section{Rotating the Hamiltonian}\label{app:ex1_rot_nl}
\label{app:rotation}

Let us recall the unitary transformation recapped 
\begin{equation}
\begin{split}
\begin{cases}
p_n = \frac{a_n+b_n}{\sqrt{2}}\\
f_n = \frac{a_n-b_n}{\sqrt{2}} \\
c_n=c_n
\end{cases}
 \Leftrightarrow\qquad
\begin{cases}
a_n = \frac{p_n+f_n}{\sqrt{2}}\\
b_n = \frac{p_n-f_n}{\sqrt{2}} \\
c_n=c_n
\end{cases}
\end{split}
\label{eq:transformation_app}
\end{equation}
In the new coordinates, the single particle Hamiltonian $H_0$ for $\phi=\pi$ reads  
\begin{equation}
\begin{split}
H_0 &=\sum_n \left[ a_n^*  (c_n+c_{n+1}) + b_n^* (c_n- c_{n+1})  +  \text{H.c.} \right] \\
&=\frac{1}{\sqrt{2}} \sum_n \left[ (p_n+f_n)^*  (c_n+c_{n+1})\right.\\
&\qquad\quad\left. + (p_n-f_n)^* (c_n- c_{n+1})  +  \text{H.c.} \right] \\
&=\frac{1}{\sqrt{2}} \sum_n \left[ p_n^*c_n+p_n^* c_{n+1}+f_n^*c_n+f_n^*c_{n+1}\right.\\
&\qquad\quad\left. + p_n^*c_n -p_n^* c_{n+1}-f_n^*c_n+f_n^*c_{n+1}\  +  \text{H.c.} \right] 
\\
&=\sqrt{2} \sum_n \left[ p_n^*c_n + f_n^*c_{n+1}+  \text{H.c.} \right] \\
&=\sqrt{2} \sum_n \left[ c_n(p_n^* + f_{n-1}^*) +  \text{H.c.} \right] 
\end{split}
\label{eq:Ham0_app}
\end{equation}
Let us expand the terms $|a_n|^4$ and $|b_n|^4$ of the interaction Hamiltonian in the new coordinates 
\begin{equation}
\begin{split}
|a_n|^4 &= \frac{1}{4} [(p_n + f_n) (p_n^* + f_n^*)]^2\\ 
&= \frac{1}{4} (|p_n|^2 + p_n^*f_n + p_nf_n^*+|f_n|^2 )^2\\
&= \frac{1}{4} (|p_n|^4 + (p_n^*f_n)^2 + (p_nf_n^*)^2+|f_n|^4 \\
&\qquad + 2 |p_n|^2p_n^*f_n + 2 |p_n|^2 p_nf_n^*+ 2 |p_n|^2 |f_n|^2  \\
&\qquad + |p_n|^2 |f_n|^2 + p_n^*f_n |f_n|^2 + p_nf_n^*|f_n|^2 ) \\
|b_n|^4 &= \frac{1}{4} [(p_n - f_n) (p_n^* - f_n^*)]^2\\ 
&= \frac{1}{4} (|p_n|^2 - p_n^*f_n - p_nf_n^*+|f_n|^2 )^2\\
&= \frac{1}{4} (|p_n|^4 + (p_n^*f_n)^2 + (p_nf_n^*)^2+|f_n|^4 \\
&\qquad - 2 |p_n|^2p_n^*f_n - 2 |p_n|^2 p_nf_n^*- 2 |p_n|^2 |f_n|^2  \\
&\qquad - |p_n|^2 |f_n|^2 - p_n^*f_n |f_n|^2 - p_nf_n^*|f_n|^2 ) 
\end{split}
\label{eq:Ham_nl1_a_b_app}
\end{equation}
The sum of these two expanded terms $|a_n|^4 + |b_n|^4 $ in Eq.~\eqref{eq:Ham_nl1_a_b_app} cancels many addends. 
This yields a rotated form of the interaction Hamiltonian
\begin{equation}
\begin{split}
H_1&=\frac{\gamma}{2}\sum_n \left[ |a_n|^4 + |b_n|^4 + |c_n|^4 \right]\\
&= \frac{\gamma}{4} \sum_n \left[  |p_n|^4 + (p_n^*f_n)^2 + (p_nf_n^*)^2 \right. \\ 
&\qquad\quad \left.+|f_n|^4+4|p_n|^2|f_n|^2 + 2 |c_n|^4 \right] 
\end{split}
\label{eq:Ham_nl1_app_2}
\end{equation}

\section{Thermodynamics of the nonlinear system}
\label{app:thermo}

In this section, we detail the calculation of the ground state line and the infinite temperature line, respectively obtained for $\beta=\infty$ and $\beta=0$ in the probability distribution 
\begin{equation}
\mathcal{P} = \frac{1}{Z}\exp[-\beta (H-\mu A)]
\label{eq:distro_app}
\end{equation}
where $Z$ is the the grand canonical partition function 
\begin{equation}
Z = \int \prod_{n=1}^N  d\psi_n d\psi_n^* \exp[-\beta (H-\mu A)]
\label{eq:part_funct_app}
\end{equation}
defined for the Hamiltonian $H$, the total norm $A= \sum_{n=1}^N [|a_n|^2 + |b_n|^2 + |c_n|^2]$, the inverse temperature $\beta = \frac{1}{T}$ and the chemical potential $\mu$.

\subsection{The $\beta = \infty$ line} 
\label{app:thermo_beta_inf}

Let us compute the ground state line for the nonlinear Aharonov-Bohm lattice. 
To do this, we consider the following ansatz 
\begin{equation}
\begin{split}
p_{n} = \sqrt{P}e^{i\phi_n}
\quad
c_{n} = \sqrt{C}e^{i\theta_n}
\quad
f_{n} = \sqrt{F}e^{i\rho_n}
\end{split}
\label{eq:ansatz_app}
\end{equation}
defined for the amplitudes $P,C, F$ with $\Gamma= P + F + C$, and the phases $\phi_n, \theta_n, \rho_n$. 
This results in a norm density $a = \frac{A}{3N} = \frac{\Gamma}{3}$, since 
\begin{equation}
\begin{split}
A = \sum_{n=1}^N |p_n|^2 + |c_n|^2 + |f_n|^2   = L (P + F + C) = N\Gamma 
\end{split}   
\label{eq:norm_app}
\end{equation}

\begin{figure}[h!]
    \centering
   \includegraphics[width=0.95\columnwidth]{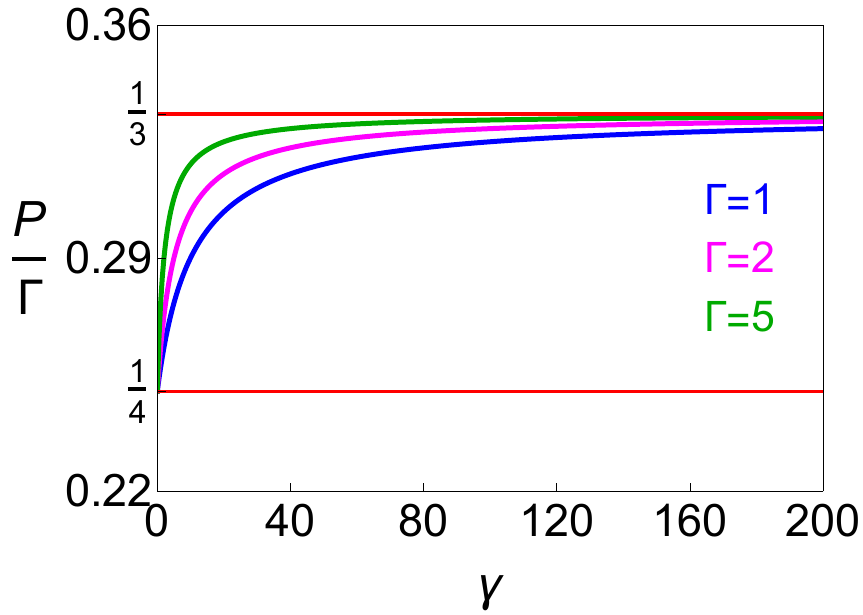}
    \caption{Positive real root $P$ of the polynomial $f(P,\Gamma,\gamma)$ rescaled by constant $\Gamma$ as function the nonlinear strength $\gamma$ for $\Gamma=1$ (blue), $\Gamma=2$ (magenta), and $\Gamma=5$ (green). }
    \label{fig:app1}
\end{figure}

\noindent
The rotated Hamiltonian -- here recapped
\begin{equation}
\begin{split}
H 
&=\sqrt{2}\sum_{n=1}^N \left[ c_n (p_n^* + f_{n-1}^*) +  \text{H.c.} \right] \\
&+ \frac{\gamma}{4} \sum_{n=1}^N \left[  |p_n|^4 + (p_n^*f_n)^2 + (p_nf_n^*)^2 \right. \\ 
&\qquad\quad \left.+|f_n|^4+4|p_n|^2|f_n|^2 + 2 |c_n|^4 \right] 
\end{split}
\label{eq:Ham_rot1_app}
\end{equation}
turns to 
\begin{equation}
\begin{split}
H &= 2 \sqrt{2}\sum_{n=1}^N \sqrt{ C } \left[ \sqrt{P}  \cos (\phi_n - \theta_n)  +   \sqrt{F} \cos (\rho_{n-1} - \theta_n) \right]\\
&+\frac{\gamma}{4}\sum_{n=1}^N \left[ P^2 + 2 P F (2+ \cos 2(\phi_n - \rho_n)) +F^2   + 2C^2 \right] \\
\end{split}
\label{eq:Ham_ph_amp_app}
\end{equation}
To minimize this Hamiltonian, we assume the following phase relations which ensure all $\cos$ terms to be $-1$
\begin{equation}
\begin{split}
\begin{cases}
\phi_n - \theta_n = \pi \\
\rho_{n-1} - \theta_n = \pi  \\
2(\phi_n - \rho_n) = \pi 
\end{cases}
\,\Leftrightarrow \qquad
& \begin{cases}
\phi_n  = \theta_n + \pi \\
 \rho_n = \phi_n  + \frac{\pi}{2}\\
 \theta_n = \rho_{n-1} + \pi  \\
\end{cases} 
\end{split}
\label{eq:Ham_lin_app}
\end{equation}
For $C = \Gamma - P - F$, the phase relations yield the energy density $h=\frac{H}{3N}$
\begin{equation}
\begin{split}
h &=\frac{H}{3N} = \frac{H_0}{3N} + \frac{\gamma}{2} \frac{H_1}{3N} \\ 
&=  - \frac{2}{3} \sqrt{2( \Gamma - P - F) } \left[ \sqrt{P}  +   \sqrt{F} \right]   \\
&+ \frac{\gamma}{12} 
\left[ \frac{1}{2} [P + F]^2   + (\Gamma - P - F)^2 \right] 
\end{split}
\label{eq:Ham_lin_app2}
\end{equation}
\noindent
The partial derivatives of $h$ over $P$ and $F$ respectively are 
\begin{equation}
\small
\begin{split}
\frac{\partial h}{\partial P} & =\frac{1}{6}\left\{ - 2 \sqrt{2} \frac{2 P + F + \sqrt{PF } -\Gamma }{\sqrt{P(\Gamma-P-F)} } +\frac{\gamma}{2} (3P + 3F - 2\Gamma) \right\}\\
\frac{\partial h}{\partial F} & =\frac{1}{6}\left\{ - 2 \sqrt{2} \frac{2 F + P + \sqrt{PF } -\Gamma }{\sqrt{F(\Gamma-P-F)} } +\frac{\gamma}{2} (3P + 3F - 2\Gamma)  \right\}
\end{split}
\label{eq:Ham_lin_app3}
\end{equation}
Setting the minimum implies search for $\frac{\partial h}{\partial P}=0=\frac{\partial h}{\partial F}$ -- an identity that implies that $P=F$. 
Following this, the energy density $h$ and its derivative $\frac{\partial h}{\partial P}$ in Eqs.~(\ref{eq:Ham_lin_app2},\ref{eq:Ham_lin_app3}) reduce to 
\begin{equation}
\begin{split}
h &=  - \frac{4}{3} \sqrt{2P( \Gamma - 2P)  } 
+ \frac{\gamma}{4} \frac{1}{3} 
\left[  2P^2   + (\Gamma - 2P)^2 \right]  \\
\frac{\partial h}{\partial P} & =\frac{2}{3}\left\{  - \sqrt{2} \frac{\Gamma - 4 P }{\sqrt{P(\Gamma-2 P )} } + \frac{\gamma}{2} (3P - \Gamma) \right\}
\end{split}
\label{eq:Ham_lin_app4}
\end{equation}
Setting the condition for the minumum $\frac{\partial h}{\partial P} =0$ implies
\begin{equation}
\small 
\begin{split}
&   \sqrt{2} (\Gamma - 4P )  = \frac{\gamma}{2} (3P - \Gamma) \sqrt{P(\Gamma-2 P )}\\
    &
  2 \Gamma^2 + 32 P^2 - 16 \Gamma P
  = \frac{\gamma^2}{4}   ( P \Gamma^3 - 8P^2 \Gamma^2 +21 P^2 \Gamma - 18P^4) 
\end{split}
\label{eq:Ham_lin_app5}
\end{equation}
which ultimately yields a $4^{th}$-order polynomial in $P$ 
\begin{equation}
\begin{split}
f(P,\Gamma,\gamma) &=  18\frac{\gamma^2}{4} P^4 - 21 \frac{\gamma^2}{4}  \Gamma P^3 + 8 (4 + \frac{\gamma^2}{4} \Gamma^2) P^2 \\
& - (16 \Gamma  + \frac{\gamma^2}{4} \Gamma^3) P  + 2 \Gamma^2 
\end{split}
\label{eq:poly_app}
\end{equation}
for given norm density $a = \frac{\Gamma}{3}$ and nonlinear strength $\gamma$. 
Among the four roots of the polynomial $f$ in Eq.~\eqref{eq:poly_app} for given $\gamma,\Gamma$, only one is real and positive. This root, for a given pair $\{\gamma,\Gamma\}$, defines the profile of the ground state in Eq.~\eqref{eq:ansatz_app} with the phase relations in Eq.~\eqref{eq:Ham_lin_app} and its energy.  \\
\\
We plot the numerically computed root as function of the nonlinear strength $\gamma$ in Fig.~\ref{fig:app1} for three different values $\Gamma=1,2,5$. 
We observe how, in all three cases, the rescaled root $\frac{P}{\Gamma}$ ranges from $\frac{P}{\Gamma}=\frac{1}{4}$ for $\gamma=0$ (linear limit) to $\frac{P}{\Gamma}=\frac{1}{3}$ for $\gamma\longmapsto +\infty$ (purely nonlinear limit).

\subsection{The $\beta = 0$ line}
\label{app:thermo_beta_0}

In the infinite temperature limit  $\beta=0$, the probability distribution $\mathcal{P}$ in Eq.~\eqref{eq:distro_app} reduces to $\mathcal{P}\sim \exp(m A)$ where $m=\beta\mu$ is supposed to be finite and negative -- {\it i.e.} $-\infty <m<0$. Notice that, since $A$ does not depend on the phases of the variables $a_n, b_n, c_n$,  the partition function $P$ is uniform in phases. 
For our convenience, let us rewrite the distribution as $\mathcal{P}\sim \exp(-\alpha A)$ for positive $\alpha = - m = -\beta\mu$.

Let us compute the average norm and energy densities in this limit, specifically 
\begin{equation}
a= \frac{\langle A \rangle }{3N}
\qquad \qquad 
h= \frac{\langle H \rangle }{3N}
\end{equation}
where $\langle \cdot \rangle$ is the average over $P$.
It is convenient to express $P$ as a function of local masses $x_n=|a_n|^2$, $y_n=|b_n|^2$, $z_n=|c_n|^2$. This yields 
\begin{equation}
\mathcal{P} =  \alpha^{3N} \exp[-\alpha \sum_{n=1}^N (x_n+ y_n +z_n)] 
\label{eq:P_app}
\end{equation}
The average norm density $a$ reads
\begin{equation}
a = \frac{1}{3N} (3N)\left( \alpha \int_0^\infty dx \exp(- \alpha x) x \right) = \frac{1}{\alpha} 
\label{eq:P_app2}
\end{equation}
which implies that $\alpha= \frac{1}{a}$.\\
\\
Since all hopping terms of $H_0$ vanish due to the flat distribution of phases, the average energy density $h$ reads 
\begin{equation}
h= \frac{1}{3N} (3N) \left(  \frac{\alpha\gamma}{2} \int_0^\infty dx \exp(-\alpha x) x^2 \right) = \frac{\gamma}{\alpha^2}
\label{eq:P_app3}
\end{equation}
Therefore, for $\beta=0$ we obtain the relation $h= \gamma a^2$.

\section{Details on numerical simulations}
\label{app:num_sim}

To sample equilibrium and nonequilibrium thermodynamic states, lattice equations~(\ref{eq:Ham_eq_Nlin}) have been numerically integrated 
according to a standard 4th-order Runge-Kutta algorithm~\cite{press2007numerical} with a time step not larger than $10^{-3}$ units.
In addition to the Hamiltonian bulk dynamics,
thermal boundary reservoir have been implemented according to stochastic Monte Carlo rules, see next paragraph. Averages of observables in stationary states
have been computed for time periods of at least $5\times 10^6$ and up to $5\times10^7$ units after suitable relaxation transients of at least $10^5$ units.

\paragraph{\it Definition of heat baths at the chain edges --}
    
    Grand-canonical Monte Carlo reservoirs are implemented at the chain edges according to~\cite{iubini12}. Here we focus on the left reservoir,
    imposing an inverse temperature $\beta_L$ and chemical potential $\mu_L$. An analogous definition applies to the reservoir at the right edge.
     
    The reservoir introduces stochastic perturbations of the field variables $\psi_n=(a_n,b_n,c_n)$ at lattice site $n=1$, namely $\psi_1 \to \psi_1+ \delta \psi_1$,
    where $\delta\psi_1=(\delta p_1 + i\delta q_1)/\sqrt{2}$. In between two successive stochastic moves, the evolution of the system is deterministic
    and governed by the Hamiltonian equations in~(\ref{eq:Ham_eq_Nlin}). Moves are implemented with a time period $t_b$.
    Perturbed variables are accepted or rejected according to a Metropolis cost function 
    $\exp\{-\beta_L (\Delta H - \mu\Delta A)\}$, where $\Delta H$ and $\Delta A$ are respectively the energy  and norm variations produced by the
    perturbation $\delta \psi_1$. Perturbations $\delta p_1$ and  $\delta q_1$ are independent random variables  extracted from an uniform distribution
    in the interval $[-\epsilon, \epsilon]$. The parameters $\epsilon$ and $t_b$  determine effective system-bath coupling.
    
    We have verified that different prescriptions on the bath dynamics, as for example a Poissonian distribution of interaction times in place of the
    constant-pace rule here adopted do not significantly modify the statistical properties of the model. Simulations are performed with $\epsilon=2$
    and $t_b=0.1$ time units. We have verified that this choice of parameters allows to minimize boundary resistances in the whole range of parameter
    explored in this study.

     \paragraph{\it Definition of norm and energy currents --}

Currents are measured at the edge of the system by monitoring the exchanges of norm and energy between the system and the bath.
For the left edge,
\begin{equation}
\begin{split}
    j_a^{(L)}& = \frac{1}{\tau} \sum_{t_k< \tau} \delta a_1(t_k)\\
    j_h^{(L)}& = \frac{1}{\tau} \sum_{t_k< \tau} \delta h_1(t_k)
\end{split}   
\label{eq:ja_jh}
\end{equation}
where $ \delta a_1(t_k)$ and $\delta h_1(t_k)$ are respectively the norm and energy variations at the first lattice site produced by a reservoir update occurring at time $t_k$. An average over a sufficiently long time $\tau\gg 1$ is performed.  An analogous expression for the currents at the right edge, $-j_a^{(R)}$ and 
$-j_h^{(R)}$ can be obtained form Eq.~(\ref{eq:ja_jh}) upon replacing $a_1(t_k) \to a_N(t_k)$ and $h_1(t_k) \to h_N(t_k)$. 

\paragraph{\it Onsager matrix elements--}

For the numerical estimation of the Onsager coefficients $L_{xy}$ in Eq.~(\ref{eq.Ons2}) the following procedure is employed, see~\cite{iubini12,iubini23} for details. A reference working point $(\beta,m)$ is fixed  and two independent simulations are performed imposing either a pure temperature gradient $\beta_y$ (with $m_y=0$) or a pure `chemical' gradient $m_y$ (with $\beta_y=0$). From the measure of stationary currents $j_a$ and $j_j$ one obtains
\begin{equation}
    \begin{split}
        L_{ah}&=j_a/\beta_y   \quad(m_y=0)\\
        L_{hh}&=j_h/\beta_y    \quad(m_y=0)\\
        L_{aa}&=-j_a/m_y \quad(\beta_y=0)\\
        L_{ha}&=-j_h/m_y  \quad(\beta_y=0)
    \end{split}
\end{equation}
Nonequilibrium simulations are performed on a chain with $N=100$ evolved for $5\times 10^7$ units.

\section{Equilibrium thermodynamics of the linear chain $H_0$ }
\label{app:thermo_lin}

In general, for a lattice with three bands $\omega_i(k)$ with $i=0,\pm$ represented with normal modes $\varphi_i(k)$, the norm and Hamiltonian energy read 
\begin{eqnarray}
\small
\label{eq:normal_modes_app}
A&=&\sum_{k} |\varphi_0(k)|^2 + |\varphi_+(k)|^2 + |\varphi_{-}(k)|^2 \\
H&=&\sum_{k} \omega_0(k) |\varphi_0(k)|^2 + \omega_+ (k) |\varphi_+(k)|^2 + \omega_{-}(k) |\varphi_{-}(k)|^2  \notag 
\end{eqnarray}
The partition function $Z$ in Eq.~\eqref{eq:part_funct_app} 
\begin{equation}
    Z= \int \prod_k \prod_{i=0,\pm} d \varphi_i(k) \,d \varphi_i^*(k) e^{-\beta H +m A}
\end{equation}
expressed in the normal modes for $m = \beta\mu$ reduces to a product of Gaussian integrals 
\begin{equation}
\small
\begin{split}
    &Z 
    =\prod_{k}\prod_{i=0,\pm } \frac{2\pi}{(\beta \omega_i(k) -m)} \\
    =& (2\pi)^{3N} \left( \frac{1}{\beta \omega_0-m}\right)^N\left( \frac{1}{\beta \omega_+-m}\right)^N\left( \frac{1}{\beta \omega_{-}-m}\right)^N
\end{split} 
\label{eq:part_H0}
\end{equation} 
The average norm density is
\begin{equation}
\begin{split}
    a &= \frac{1}{3N} \frac{\partial\log Z}{\partial m} = \frac{1}{3N} \frac{\sum_{k,i} \log(\frac{2\pi}{\beta\omega_i(k) -m})}{\partial m}\\
    &=\frac{1}{3N}\sum_{k,i} \frac{1}{\beta\omega_i(k)-m}
\end{split}    
\label{eq:RJ}
\end{equation}
Let us recall that in our system, the central band is in general independent on the wave vector $k$ -- {\it i.e.} $\omega_0=0$ -- and the two dispersive bands are symmetric -- {\it i.e.} $\omega_+(k) = -\omega_{-}(k)$ for all $\phi$. 
This implies that
\begin{equation}
\begin{split}
    \frac{1}{\beta\omega_+(k)-m} + \frac{1}{\beta\omega_{-}(k)-m} = 
    \frac{2m }{\beta^2\omega^2_+(k)-m^2}
\end{split}
\label{eq:razion_app}
\end{equation}
Then, Eq.\eqref{eq:RJ} reduces to
\begin{equation}
\small
  a= \frac{1}{3}\left[\frac{1}{2\pi} \int_0^{2\pi} \frac{2 m }{4\beta^2(1+\cos(\phi/2)\cos(k)) - m^2}\, dk  -\frac{1}{m} \right]
  \label{eq:a_ave_H0_app}
\end{equation}
Setting $q=[m/(2\beta)]^2=\mu^2/4$ yields the solution of the integral 
\begin{equation}
 \int_0^{2\pi} \frac{1 }{(1+\cos(\phi/2)\cos(k)) - q}\, dk =  -\frac{2\pi}{\sqrt{K-2q+q^2}}
\label{eq:gen_int}
\end{equation}
where we set $q>1+\sqrt{1-K}$ with $K=1/2(1-\cos \phi)$ to get real solutions. 
This implies the relation 
\begin{equation}
    a=-\frac{1}{3m}\left( \frac{2q}{\sqrt{K-2q+q^2}}  +1\right) 
    \label{eq:a_gen_app}
\end{equation}
The condition $q>1+\sqrt{1-K}$ is equivalent to  $|\mu|>2\sqrt{1+\cos(\phi/2)}$.
Furthermore, along a similar approach, it follows that 
\begin{equation} 
\begin{split}
    h&=-\frac{1}{3N} \frac{\partial\log Z}{\partial \beta} = \frac{1}{3N}\sum_{k,i} \frac{\omega_i(k)}{\beta\omega_i(k)-m} \\
    &= \frac{1}{3N\beta}\sum_{k,i} \frac{\beta\omega_i(k)-m+m}{\beta\omega_i(k)-m} = \frac{1}{\beta}+\mu a
\end{split}
\label{eq:a_h_H0_app}
\end{equation}
{\it i.e.} the relation between the norm and the energy density is $h=\frac{1}{\beta}+\mu a$.

\section{Landauer theory for linear transport}
\label{app:Landauer_lin}

In the linear limit $\gamma=0$, we employ Landauer theory to describe stationary transport~\cite{sheng1996}. 
The stationary norm current $j_a$ and energy current $j_h$ measured at the edges of the chain connected to the baths are defined as
\begin{eqnarray}
    \label{eq:Landauer_int_ja_app}
     j_a &= \int_{\mathcal{D}_\omega} d\omega\, t(\omega)[f_L(\omega)-f_R(\omega)]\\ 
     \label{eq:Landauer_int_jh_app}
     j_h  &= \int_{\mathcal{D}_\omega} d\omega\,\omega t(\omega)[f_L(\omega)-f_R(\omega)]
\end{eqnarray} 
where $f_{L,R}$ is the Rayleigh-Jeans (RJ) distribution, $f_{L,R}=\frac{1}{\beta(\omega-\mu)}$, $\mathcal{D}_\omega$ is the available frequency spectrum, and $t(\omega)$ is the transmission coefficient. 
The specific form of $t(\omega)$ depends on the system-bath coupling, and in this case we assume it constant $t = 1$.

The norm flux $j_a$ in Eq.~\eqref{eq:Landauer_int_ja_app} reads 
\begin{equation}
\begin{split}
j_a &= \Phi(\beta_L,\mu_L,\phi) - \Phi(\beta_R,\mu_R,\phi) \\
\end{split} 
\label{eq:ja_int_app}
\end{equation}
where 
\begin{equation}
\begin{split}
  \Phi(\beta,\mu,\phi)  &= \frac{1}{\beta} \int_{\mathcal{D}_\omega} d\omega\, \frac{1}{\omega-\mu}
\end{split} 
\label{eq:ja_int_app_2}  
\end{equation}
The available frequency spectrum $\mathcal{D}_\omega$ is the union of the three bands: 
two dispersive $\omega_{+}\in [\omega_l,  \omega_u]$ and $\omega_{-}\in [- \omega_u, - \omega_l]$  for $\omega_l=2\sqrt{1-\cos{\phi/2}}$ and $\omega_u=2\sqrt{1+\cos{\phi/2}}$, and a flat band $\omega_0=0$. Over these two symmetric intervals, the integral of RJ distribution reads
\begin{equation}
\begin{split}
  \Phi(\beta,\mu,\phi)  
  &= \frac{1}{\beta}\log\left[ \frac{(\omega_u-\mu)}{(\omega_u+\mu)} \frac{(\omega_l+\mu)}{(\omega_l-\mu)} \right]
\end{split} 
\label{eq:ja_int_app_3}  
\end{equation}
The expression for $j_a$ in Eq.~(\ref{eq:ja_int_app}) for $\mu_L = \mu_R\equiv \mu$ then reduces to
\begin{equation}
\begin{split}
j_a 
&= \log\left[ \frac{(\omega_u-\mu)}{(\omega_u+\mu)} \frac{(\omega_l+\mu)}{(\omega_l-\mu)} \right]
\left[\frac{1}{\beta_L} - \frac{1}{\beta_R} \right]
\end{split} 
\label{eq:ja_int_app_4}
\end{equation}
The energy flux $j_h$ in Eq.~\eqref{eq:Landauer_int_jh_app} reads 
\begin{equation}
    j_h = \Lambda(\beta_L,\mu_L,\phi) - \Lambda(\beta_R,\mu_R,\phi)
\label{eq:jh_int_app}     
\end{equation}
where
\begin{equation}
\begin{split}
  \Lambda(\beta,\mu,\phi)  
  &= \frac{1}{\beta} \int_{\mathcal{D}_\omega} d\omega\, \frac{\omega}{\omega-\mu}\\
 & = \frac{2}{\beta}(\omega_u - \omega_l) + \mu\Phi(\beta,\mu,\phi)
\end{split}
\label{eq:jh_int_app_2}     
\end{equation}
For $\mu_L = \mu_R\equiv \mu$,   Eq.~\eqref{eq:jh_int_app} reduces to
\begin{equation}
\small
\begin{split}
j_h 
&= \left\{ 2(\omega_u - \omega_l) + \mu \log\left[ \frac{(\omega_u-\mu)}{(\omega_u+\mu)} \frac{(\omega_l+\mu)}{(\omega_l-\mu)} \right] \right\} \left[\frac{1}{\beta_L} - \frac{1}{\beta_R} \right]
\end{split}    
\label{eq:jh_int_app_3}     
\end{equation}
Ultimately, the fluxes $j_a$ and $j_h$ can be written as 
\begin{equation}
\begin{split}
j_a &= C_\phi
\left[\frac{1}{\beta_L} - \frac{1}{\beta_R} \right]\\
j_h &= \left\{ 2(\omega_u - \omega_l) + \mu C_\phi \right\} \left[\frac{1}{\beta_L} - \frac{1}{\beta_R} \right]
\end{split}    
\label{eq:ja_jh_app}     
\end{equation}
for the coefficient $C_\phi = \log\left[ \frac{(\omega_u-\mu)}{(\omega_u+\mu)} \frac{(\omega_l+\mu)}{(\omega_l-\mu)} \right]$. 
Notice that, for $\phi=\pi$ where all bands turn flat, since $\omega_l=\omega_u$ both fluxes $j_a$ and $j_h$ in Eqs.~(\ref{eq:ja_int_app},\ref{eq:jh_int_app}) vanish for any choices of $\{\beta_L,\mu_L\}$ and $\{\beta_R,\mu_R\}$.

\section{Size dependence of the fluxes}
\label{app:systemsize}

\begin{figure}[h!]
    \centering
   \includegraphics[width=0.95\columnwidth]{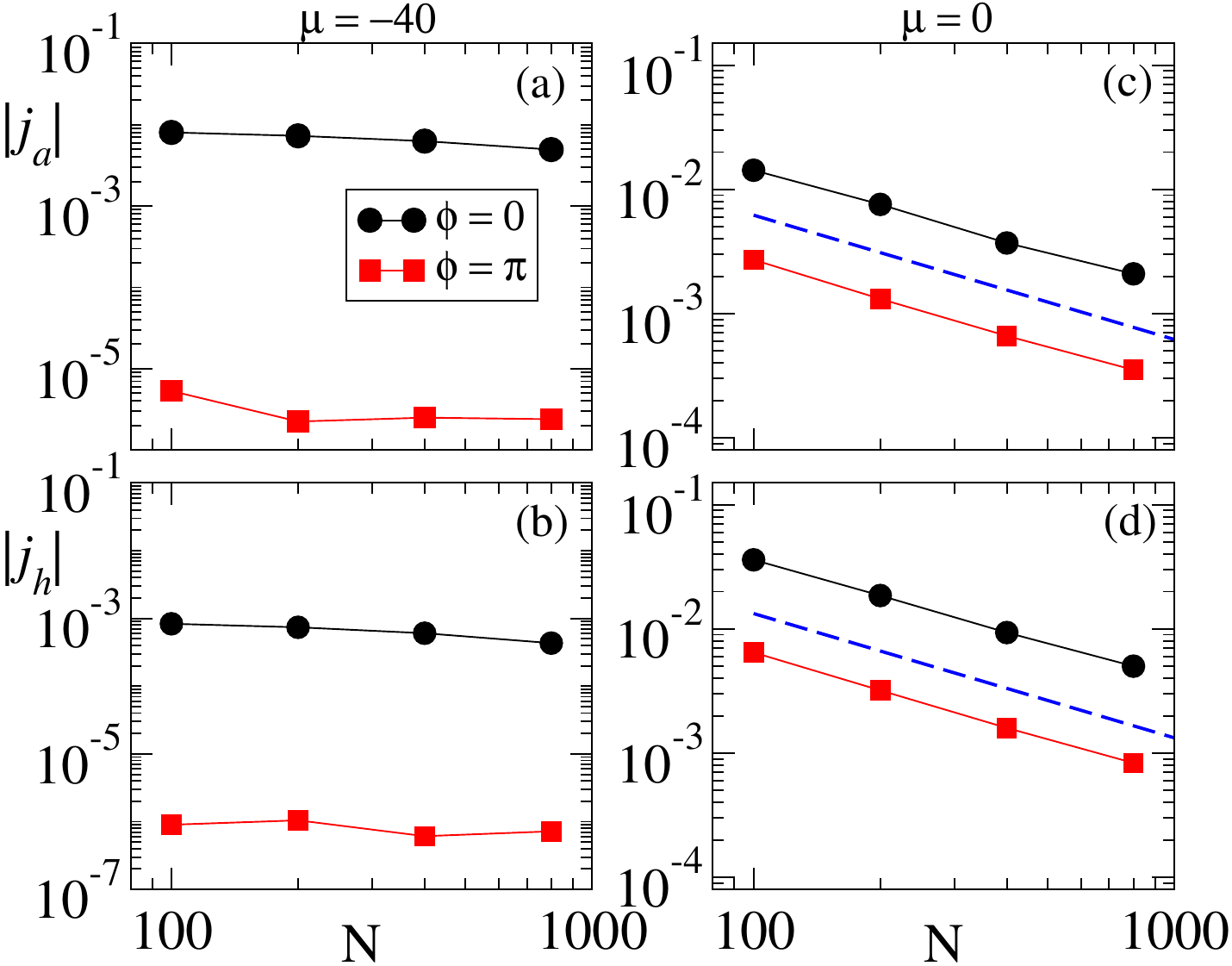}
    \caption{
    Size dependence of stationary absolute fluxes of norm (first row) and energy (second row) in response to a temperature gradient, $\beta_L=1$, $\beta_R=0.5$  for $\gamma = 2$. 
    Left column shows the low-norm (quasi-linear) regime with $\mu=-40$.
    Right column refers to $\mu=0$ with fully developed nonlinearity. Blue dashed lines refer to the diffusive scaling $j\sim N^{-1}$. 
    }
    \label{fig:app2}
\end{figure}

In Fig.~\ref{fig:app2} we report the size dependence of currents in the weakly nonlinear regime (panels (a) (b)) and in the full nonlinear one (panles (c) and(d)).
In this study, nonlinearity was kept fixed $(\gamma=2)$ while we numerically explored via Monte-Carlo nonequilibrium simulations two different regions of the phase diagram in Fig.~\ref{fig:2} characterized by $\mu=-40$ (weakly nonlinear) and $\mu=0$ (full nonlinear). For simplicity,  we fixed the same temperature gradient $\beta_L=1$, $\beta_R=0.5$  and vanishing $\mu$ gradient for both regimes. The two above points correspond to norms $a\simeq 0.04$ $(\mu=-40)$ and 
$a=1.1$ $(\mu=0)$. 
In the weakly nonlinear regime, stationary currents display a very weak dependence on the system size, although strong finite-size effects are expected~\cite{lepri2003thermal}. A comparison between $\phi=0$ (black dots) and $\phi=\pi$ (red squares) displays a difference of around three orders
of magnitude in the corresponding currents, in agreement with the caging effect observed in the linear limit. Accordingly, for for $\phi=0$ the system
behaves as a ballistic conductor, while for $\phi=\pi$ it is almost an insulator.
In the nonlinear regime diffusive transport is observed, with both currents scaling as $1/N$ and a reduced, still not negligible, dependence on the flux $\phi$. 

\end{document}